\documentclass[runningheads]{llncs}
\usepackage[T1]{fontenc}
\usepackage{tabularx}
\usepackage{adjustbox}
\usepackage{booktabs}
\usepackage{subcaption}
\usepackage{booktabs}
\usepackage{graphicx}
\usepackage{amsmath}
\usepackage{amsfonts}
\usepackage{array}
\usepackage{float}
\usepackage{hhline}
\usepackage[dvipsnames]{xcolor}
\usepackage{multirow}
\usepackage{url}
\usepackage{tikz}
\usepackage{cite}
\usepackage{placeins}

\usepackage{soul}

\usepackage[colorlinks=true]{hyperref}
\begin{document}
\title{Evaluating the Quality of Brain MRI Generators}
\titlerunning{Evaluating the Quality of Brain MRI Generators}
%
\author{Jiaqi Wu $\dagger$ \inst{1} \orcidID{0009-0003-7108-079X} \and
Wei Peng $\dagger$ \inst{1}\orcidID{0000-0002-2892-5764} \and\\
Binxu Li \inst{1}\orcidID{[0009-0001-8044-591X} \and
Yu Zhang \inst{2}\orcidID{0000-0001-9726-6400}\and \\
Kilian M. Pohl\inst{1}\thanks{Corresponding Author: kpohl@stanford.edu; \(\dagger\) indicates equal contribution}\orcidID{0000-0001-5416-5159}}

\authorrunning{Wu et al.}
\institute{{\small Stanford University, Stanford, CA 94305
\and
 Lehigh University, Bethlehem, PA 18015  
}}


\maketitle              
%
\begin{abstract}
Deep learning models generating structural brain MRIs have the potential to significantly accelerate discovery of neuroscience studies. However, their use has been limited in part by the way their quality is evaluated. Most evaluations of generative models focus on metrics originally designed for natural images (such as structural similarity index and Fr\'echet inception distance). As we show in a  comparison of 6 state-of-the-art generative models trained and tested on over 3000 MRIs, these metrics are sensitive to the experimental setup and inadequately assess how well brain MRIs capture macrostructural properties of brain regions (i.e., anatomical plausibility). This shortcoming of the metrics results in inconclusive findings even when qualitative differences between the outputs of models are evident. We therefore propose a framework for evaluating models generating brain MRIs, which requires uniform processing of the real MRIs, standardizing the implementation of the models, and automatically segmenting the MRIs generated by the models. The segmentations are used for quantifying the plausibility of anatomy displayed in the MRIs. To ensure meaningful quantification, it is crucial that the segmentations are highly reliable. Our framework rigorously checks this reliability, a step often overlooked by prior work. Only 3 of the 6 generative models produced MRIs, of which at least 95$\%$ had highly reliable segmentations. More importantly, the assessment of each model by our framework is in line with qualitative assessments, reinforcing the validity of our approach. The code of this framework is available via
\url{https://github.com/jiaqiw01/MRIAnatEval.git}.
\end{abstract}

\section{Introduction}

\begin{figure}[!t]
    \centering
    \resizebox{\columnwidth}{!}{
    \tikzset{every picture/.style={line width=0.75pt}} 
    
    \begin{tikzpicture}[x=0.75pt,y=0.75pt,yscale=-1,xscale=1]
    
    \newcommand{\qcscore}{395}
    \draw (297,124.5) node  {\includegraphics[width=429pt,height=380.75pt]{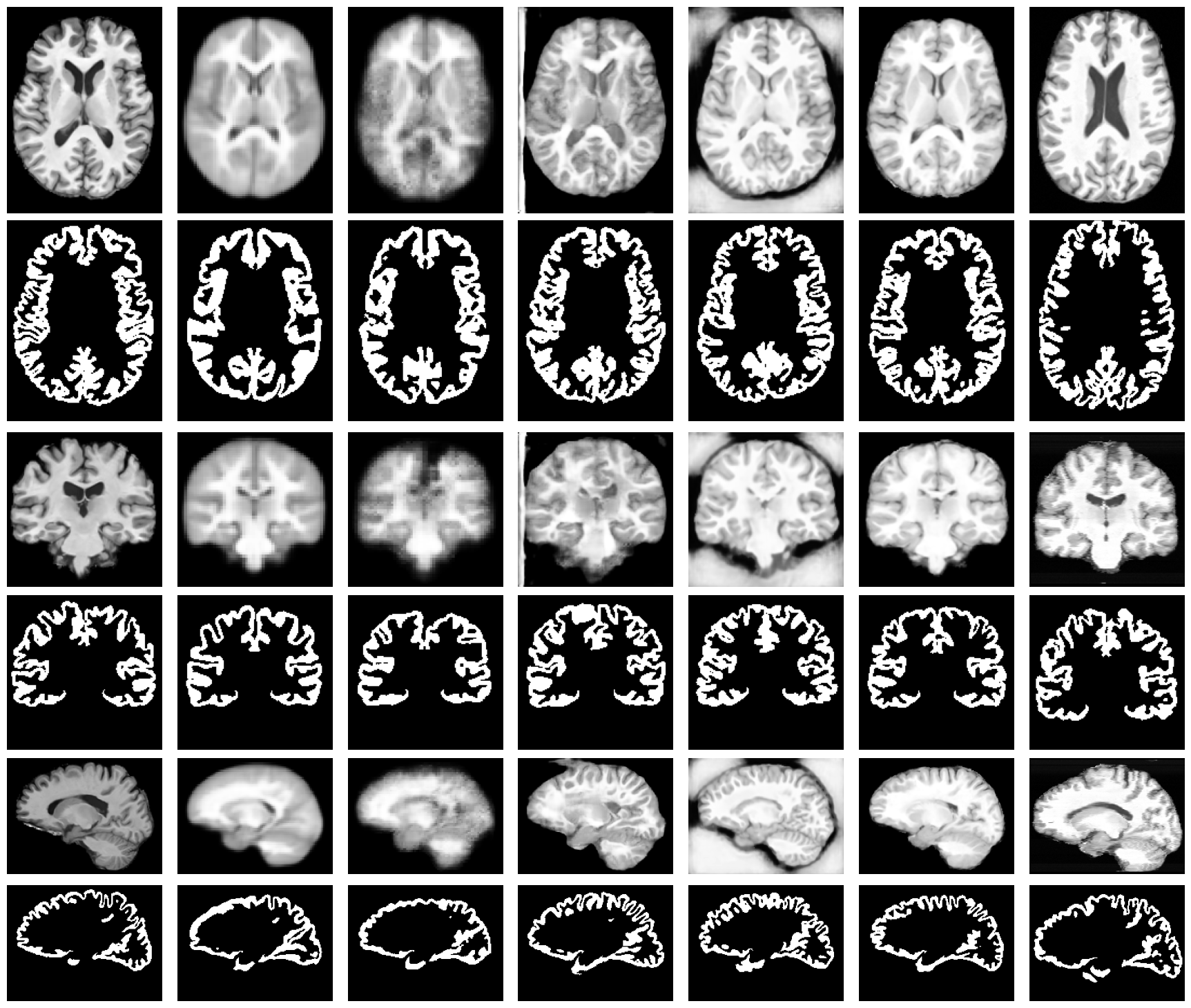}};
    \draw (40,380) node [anchor=north west][inner sep=0.75pt]  [font=\footnotesize] [align=left] {Real};
    \draw (105,380) node [anchor=north west][inner sep=0.75pt]  [font=\footnotesize] [align=left] {VAE-GAN};
    \draw (190,380) node [anchor=north west][inner sep=0.75pt]  [font=\footnotesize] [align=left] {$\alpha$-WGAN};
    \draw (265,380) node [anchor=north west][inner sep=0.75pt]  [font=\footnotesize] [align=left] {HA-GAN};
    \draw (340,380) node [anchor=north west][inner sep=0.75pt]  [font=\footnotesize] [align=left] {MONAI-LDM};
    \draw (440,380) node [anchor=north west][inner sep=0.75pt]  [font=\footnotesize] [align=left] {MedSyn};
    \draw (525,380) node [anchor=north west][inner sep=0.75pt]  [font=\footnotesize] [align=left] {cDPM};
    \draw (35,\qcscore) node [anchor=north west][inner sep=0.75pt]  [font=\footnotesize] [align=left] {0.735};
    \draw (115,\qcscore) node [anchor=north west][inner sep=0.75pt]  [font=\footnotesize] [align=left] {\textit{0.640}};
    \draw (200,\qcscore) node [anchor=north west][inner sep=0.75pt]  [font=\footnotesize] [align=left] {\textit{0.647}};
    \draw (275,\qcscore) node [anchor=north west][inner sep=0.75pt]  [font=\footnotesize] [align=left] {0.718};
    \draw (360,\qcscore) node [anchor=north west][inner sep=0.75pt]  [font=\footnotesize] [align=left] {\textit{0.618}};
    \draw (445,\qcscore) node [anchor=north west][inner sep=0.75pt]  [font=\footnotesize] [align=left] {0.715};
    \draw (525,\qcscore) node [anchor=north west][inner sep=0.75pt]  [font=\footnotesize] [align=left] {0.738};
    \end{tikzpicture}}
    \caption{Axial, coronal, and sagittal view of real and synthetic MRIs produced by six methods. The corresponding gray matter segmentations are produced by $Synthseg^{+}$~\cite{billot_robust_2023}, which also provides QC scores (listed below method names). Even if the MRIs are of relatively low quality (such as those generated by VAE-GAN, $\alpha$-WGAN, and MONAI-LDM), the segmentations still look good. However, their QC scores (in italic) is below 0.65 indicating low reliability.}
    \label{segvisual}
\end{figure}
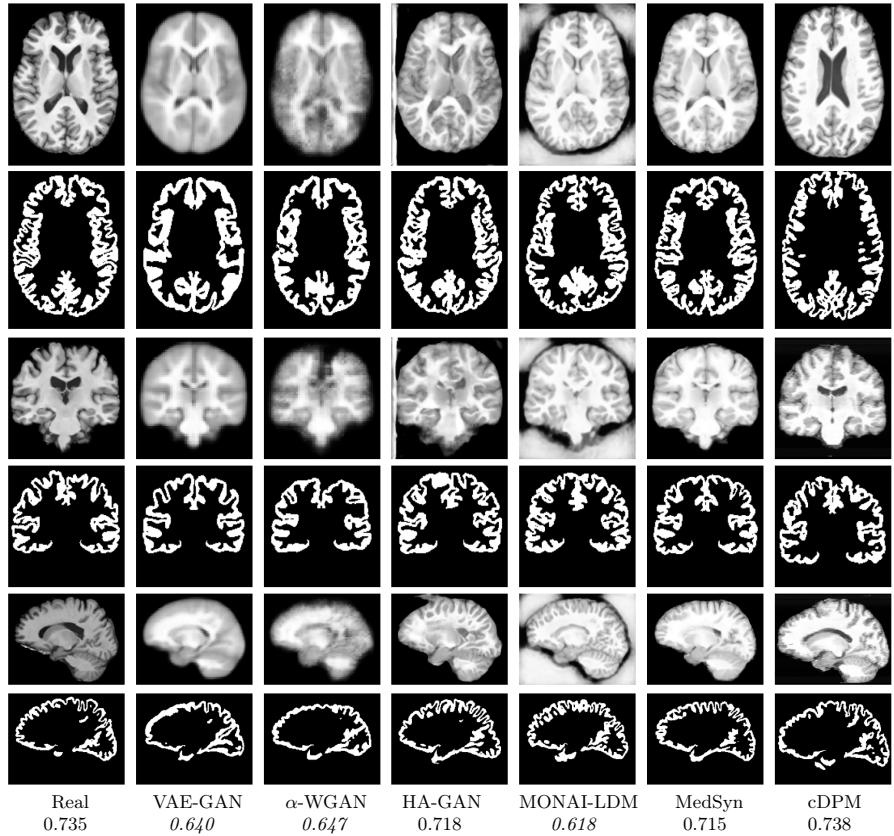

Deep learning could have a significant impact on the analysis of magnetic resonance imaging (MRI) studies for tasks such as classification \cite{ABDELAZIZISMAEL2020101779} and identification of biomarkers \cite{Bowles_Gunn_Hammers_Rueckert_2018}. Reliably training deep learning models for these tasks requires a larger number of samples~\cite{peng2022learning}, while most brain MRI studies are relatively small. Augmenting the training data with brain MRIs produced by generative models (such as shown in Fig. \ref{segvisual} ) could thus be of great value.

Current generative models, typically based on Generative Adversarial Networks (GANs)~\cite{kwon2019generation,sun2022hierarchical} or diffusion probabilistic models \cite{peng2023cDPM,pinaya2023generative,xu2023medsyn}, require a rigorous quality assessment to be useful in brain MRI studies. However, this remains an open issue as the quality of generated MRIs depends on the experimental setup (i.e., the specific implementation of the generative models and the MRIs they are trained and tested on) and the evaluation metrics. Popular metrics used for comparison are those commonly applied to 2D natural images, such as Multi-Scale Structural Similarity (MS-SSIM)~\cite{msssim}, Fr\'echet Inception Distance (FID)~\cite{heusel2017gans}, and Maximum-Mean Discrepancy (MMD)~\cite{gretton2012kernel}. However, the outcome of these metrics heavily relies on how they are applied. For example, the MMD score depends on the dimensionality (i.e., per 2D slice or 3D volume) and space (i.e., feature or image space) it is computed over \cite{peng2023cDPM, sun2022hierarchical}. For FID and MMD, the choice of feature extractor used by the generator (such as ResNet~\cite{he2016resnet} or Inception Net~\cite{szegedy2015going}) can also impact the score~\cite{heusel2017gans}. Furthermore, the FID score depends on both image quality and diversity of the generated samples making an interpretation of the score difficult \cite{NEURIPS2018_sajjadi}. Even worse, the scores might suggest that the generated MRIs are similar to real ones while the shape of brain regions shown in those MRIs is unrealistic \cite{peng2023metadataconditioned}. In \cite{peng2023metadataconditioned}, we therefore propose to quantify anatomical plausibility, i.e., how well the synthetic MRIs capture properties of brain regions.


Measuring anatomical plausibility is sensitive not only to the processing of the real MRIs and implementation of each method, but also to the automatic MRI segmentations needed for measuring properties of brain regions. To rigorously compare generative models, we therefore propose an evaluation framework that standardizes each of these three components. Novel in standardizing the automatic MRI segmentations is quantitatively assessing their reliability. Automatic segmenters, such as $Synthseg^{+}$~\cite{billot_robust_2023}, heavily rely on atlases and therefore can produce a realistic-looking label map even if the MRI is of extremely low quality (such as those produced by VAE-GAN and $\alpha$-WGAN in Fig. \ref{segvisual}). The regional measurements extracted from such a segmentation could suggest that the anatomical plausibility of the MRI is high. To avoid this scenario, our framework quantitatively checks the reliability of the label maps. If more than 5$\%$ of the segmentations from an MRI generator are deemed unreliable, we then view the quality of the MRIs created by the generative model as too low for assessment.

We use this framework to compare 3 state-of-the-art GANs and 3 diffusion methods trained and tested on over 3000 structural brain MRIs collected by three studies. Our findings reveal that metrics commonly used on 2D natural images (i.e., MS-SSIM, FID, and MMD) often yield inconclusive findings despite evident qualitative differences. In contrast, our proposed framework effectively captures these differences, providing a more accurate assessment of brain MRI generator performance.   

\section{Evaluation Framework}\label{sec:experiments}
Our framework standardizes the processing of the real MRIs, unifies the implementations of both GANs and diffusion models, and measures the anatomical plausibility of the generated MRIs. These three components are now described in further detail. 

\subsection{Standardized Processing of Real MRIs}
\label{sec:MRIProcessing} 
As in \cite{peng2023metadataconditioned}, the processing of the T1-weighted MRIs consists of denoising, bias field correction, skull stripping, intensity normalization between -1 and 1, and affine registration to the SRI atlas~\cite{rohlfing2010sri24},  which results in MRIs of with 1mm voxel resolution. We pad all MRIs to end up with 144 x 192 x 144 voxels, as some methods~\cite{xu2023medsyn} need the number of voxels to be dividedable by 3. The resulting MRIs are then used to train and test the generative models. 

\subsection{Unified Implementation of Generative Models}
\label{sec:implementation}
For a fair comparison, all generative models should be implemented using the same software platform, for which we choose PyTorch 2.0 ~\cite{paszke2019pytorch}. We then choose state-of-the-art 3D MRI generators that we can implement in PyTorch. For GANs, we select VAE-GAN~\cite{kwon2019generation}, $\alpha$-WGAN~\cite{kwon2019generation},  and HA-GAN~\cite{sun2022hierarchical}. With respect to diffusion models, we choose MONAI latent diffusion model (MONAI-LDM) \cite{pinaya2023generative}, the text-conditioned diffusion model \cite{xu2023medsyn} (MedSyn), and the conditional diffusion probabilistic model (cDPM) \cite{peng2023cDPM}. 

\subsection{Measuring Anatomical Plausibility}
\label{Two-stage} 
As in \cite{peng2023metadataconditioned}, we evaluate the anatomical plausibility of the generated MRIs by extracting regional brain measurements from them and comparing their distribution to the measurements extracted from real MRIs. In our case, each MRI is parcellated into 16 subcortical and 33 cortical regions using the automatic segmenter $Synthseg^{+}$~\cite{billot_robust_2023}. Unlike in \cite{peng2023metadataconditioned}, we check the reliability of those parcellations to ensure that the measures of anatomical plausibility can be trusted. 

We assess the reliability based on the quality control (QC) scores of 8 brain regions that $Synthseg^{+}$ provides with each segmentation (see also the example for one of the scores in Fig.  \ref{segvisual}). The reliability of an MRI is considered too low for assessment if any of the 8 QC scores are below a pre-defined threshold. We choose the threshold so that 5\% of the real MRIs of the test set fail the check. Generative models that produce MRIs whose corresponding segmentations result in an even higher failure rate are labeled as too unreliable for assessment.  


For each MRI generated by those models passing QCs, we record the volume of each brain region based on their parcellation created by $Synthseg^{+}$. From those scores, we regress out the total intracranial volume to eliminate the impact of brain size on the evaluation. For each region, the distribution of volume scores is then compared to the ones extracted from real MRI of the test set using Cohen's d score \cite{cohen2013statistical} as it was done in \cite{peng2023metadataconditioned}.

\section{Comparison of 3D MRI Generators}
Our comparison of the six generative models listed in Section \ref{sec:implementation}  is based on T1-weighted longitudinal brain MRIs from 1,236 normal controls (age range: 13 to 91 years, 589 male/647 female) pooled from three studies: the Alzheimer's Disease Neuroimaging Initiative~\cite{petersen2010alzheimer} (ADNI, 342 controls from ADNI 1, 2, 3 and GO), the National Consortium on Alcohol and Neurodevelopment in Adolescence~\cite{brown2015} (NCANDA, 621 controls from \\
NCANDA$\_$PUBLIC$\_$6Y$\_$STRUCTURAL$\_$V01~\cite{ouyang2022self}), and an in-house dataset from SRI International~\cite{zhao2021longitudinal} (SRI, 273 subjects; PI: Drs. Pefferbaum and Sullivan). 400 MRIs of 400 subjects then define the test set. They are sampled from the data set so that they uniformly span the age range of the three studies and half of them are female (see \cite{peng2023metadataconditioned} for more details). The training set consists of the remaining 836 subjects of which 3060 T1-weighted MRIs were acquired. All MRIs are processed using the pipeline described in Section \ref{sec:MRIProcessing}.  

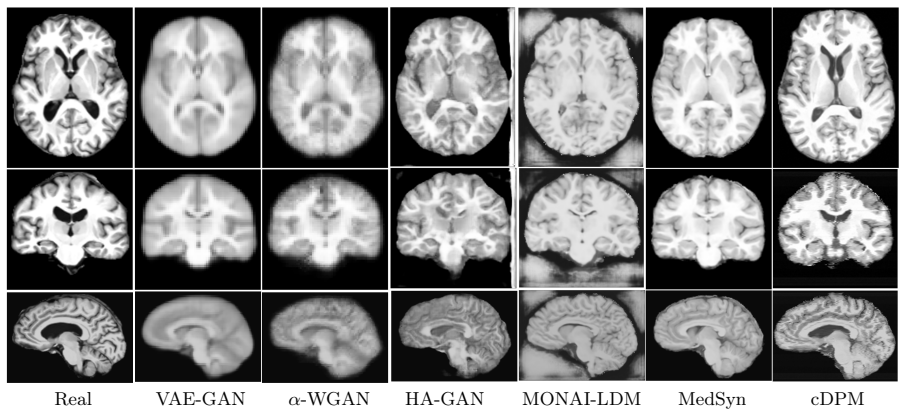
\begin{figure}[!t]
    \centering
    \resizebox{\columnwidth}{!}{
    \tikzset{every picture/.style={line width=0.75pt}} 
    
    \begin{tikzpicture}[x=0.75pt,y=0.75pt,yscale=-1,xscale=1]
    
    \draw (297,124.5) node  {\includegraphics[width=429pt,height=180.75pt]{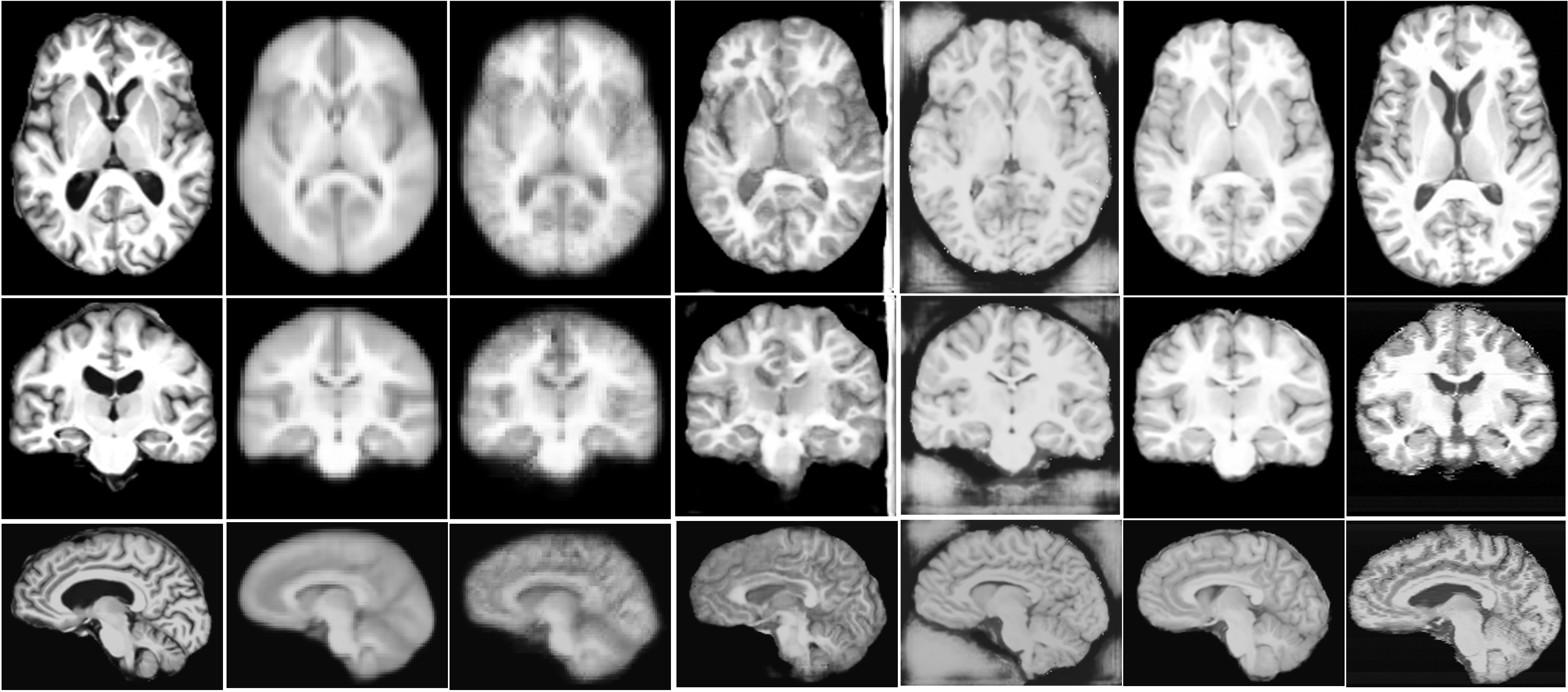}};
    
    \draw (40,249) node [anchor=north west][inner sep=0.75pt]  [font=\footnotesize] [align=left] {Real};
    \draw (105,249) node [anchor=north west][inner sep=0.75pt]  [font=\footnotesize] [align=left] {VAE-GAN};
    \draw (190,249) node [anchor=north west][inner sep=0.75pt]  [font=\footnotesize] [align=left] {$\alpha$-WGAN};
    \draw (265,249) node [anchor=north west][inner sep=0.75pt]  [font=\footnotesize] [align=left] {HA-GAN};
    \draw (340,249) node [anchor=north west][inner sep=0.75pt]  [font=\footnotesize] [align=left] {MONAI-LDM};
    \draw (440,249) node [anchor=north west][inner sep=0.75pt]  [font=\footnotesize] [align=left] {MedSyn};
    \draw (525,249) node [anchor=north west][inner sep=0.75pt]  [font=\footnotesize] [align=left] {cDPM};
    \end{tikzpicture}}
    \caption{Real MRI and synthetic MRIs from the 6 generative methods used in our comparison. In general, the MRIs from diffusion models~\cite{peng2023cDPM,xu2023medsyn} provide greater anatomical details than the GANs.}
    \label{fig:slice}
\end{figure}

After completing training, each of the six models generates 400 MRIs of which a typical example is shown in Fig.~\ref{segvisual} $\&$ \ref{fig:slice}. Visually assessing those images reveals that the outputs of VAE-GAN and $\alpha$-WGAN are blurry, the background of MRIs generated by HA-GAN and MONAI-LDM show artifacts, and those produced by the two diffusion models (MedSyn and cDPM) provide the most detail with respect to brain anatomy. In the remainder of this section, we quantitatively compare the MRIs of the six methods using metrics commonly applied to 2D natural images and 3D MRIs (Section \ref{sec:TraditionalAnalysis}). The inconclusive findings of that comparison then motivates our evaluation framework of Section \ref{sec:AnatomicalPlausibility}, which is based on anatomical plausibility (see also Section \ref{Two-stage}).

\subsection{Evaluation Based on Common Metrics }
\label{sec:TraditionalAnalysis}
As examples of metrics commonly used for assessing images, we apply FID, MMD, and MS-SSIM to the real and generated MRIs. For each model, FID and MMD compare the distributions of real and synthetic MRIs in a lower dimensional feature space. To do so, we map all MRIs into this space via a pre-trained encoder. We then document the high impact of the encoder by recoding  their scores with respect to three implementations of ResNet~\cite{he2016resnet},i.e., ResNet50 (R50), another version of ResNet50 trained on 23 datasets (R50\_23), and ResNet101 (R101)  \cite{chen2019med3d}. According to Table \ref{table:metric_table}, the lowest FID and MMD scores for VAE-GAN and $\alpha$-WGAN (the two models with the lowest quality MRIs based on visual inspection) are recorded with R101, followed by R50\_23, and  R50. Relative to the other methods, their scores for R50 were only better than MONAI-LDM. However, $\alpha$-WGAN produces the second-best scores when the encoder is R50\_23. Moreover, the two models produce the best (i.e., lowest) MMD among all the approaches when computed in the image space (Image MMD). Overall, we conclude from these results that even the worst quality images based on visually inspection can produce the best scores.  

This observation is confirmed for MS-SSIM, the only metric in this comparison independently computed from the real MRIs. Higher values are generally interpreted as better, which in this case would again point to VAE-GAN (MS-SSIM: 0.91) and $\alpha$-WGAN (MS-SSIM: 0.88) being the best approaches. However, the score of VAE-GAN is even higher than the score reported on the real MRIs (MS-SSIM: 0.88) pointing towards a lack of diversity among its MRIs and thus a mode collapse. Assuming the quality of MRIs is higher the closer the MS-SSIM score is to the one measured on the real MRI, $\alpha$-WGAN (and MedSyn) would still be the best approach. 

\begin{table}[!t]
\caption{Evaluating 400 MRIs of each approach using common metrics. Other than for MS-SSIM, lower scores are considered better with the best score being in bold and the second best underlined. For  MS-SSIM, the score closest to one recorded on the real MRIs of the test set (i.e., 0.88) is considered the best.}
\resizebox{\columnwidth}{!}{
    \begin{tabular}{lcc|cc|cc|c|c}
    \textbf{\begin{tabular}[c|]{@{}c@{}}Model\end{tabular}} & \multicolumn{2}{c|}{\textbf{\begin{tabular}[c|]{@{}c@{}}R101 \\ FID\quad MMD \end{tabular}}} & \multicolumn{2}{c|}{\textbf{\begin{tabular}[c]{@{}c@{}} R50\_23 \\ FID\quad MMD \end{tabular}}} & \multicolumn{2}{c|}{\textbf{\begin{tabular}[c]{@{}c@{}} R50 \\ FID\quad MMD \end{tabular}}}& \textbf{\begin{tabular}[c]{@{}c@{}}Image \\ MMD\end{tabular}} & \textbf{\begin{tabular}[c]{@{}c@{}}MS-\\ SSIM\end{tabular}} \\ \hline \hline
    \textbf{VAE-GAN} & 0.032   & 0.020    & 0.081 & 0.046 & 0.400 & 0.210  &  \textbf{131925} & 0.91 \\ \hline
    \textbf{$\alpha$-WGAN}  &0.032  & 0.020  & \underline{0.060} & \underline{0.040}  & 0.480 & 0.250  &  \underline{203999} & \textbf{0.88}  \\ \hline
    \textbf{HA-GAN}  & 0.035 & 0.018 &  0.079 & \underline{0.040}     & \underline{0.080} & \underline{0.043}  & 759363 & 0.78 \\ \hline
        \textbf{MONAI-LDM}  & 0.300  & 0.150   & 1.420 & 0.690  & 1.870 & 0.940 &  3314614 & 0.58  \\ \hline
    \textbf{MedSyn} &\textbf{0.012}  & \textbf{0.010}  & \textbf{0.057} & \textbf{0.037}  & \textbf{0.044}  & \textbf{0.034}  &  217897 & \textbf{0.88}  \\ \hline
    \textbf{cDPM} & \underline{0.019} & \underline{0.014}   & 0.140  & 0.082 & 0.130 & 0.081  &  586022  & 0.75  
    \end{tabular}}
\label{table:metric_table}
\end{table}

The above observations highlight the limitations of these three evaluation metrics, which are commonly used in the literature. These metrics are overly sensitive to the choice of experimental setup and do not reliably reflect the visual quality of the MRIs.

\subsection{MRI-specific Assessments}
\label{sec:AnatomicalPlausibility}
\begin{figure}[!t]
    \centering
    \resizebox{\columnwidth}{!}{
    \includegraphics[width=420pt,height=270pt]{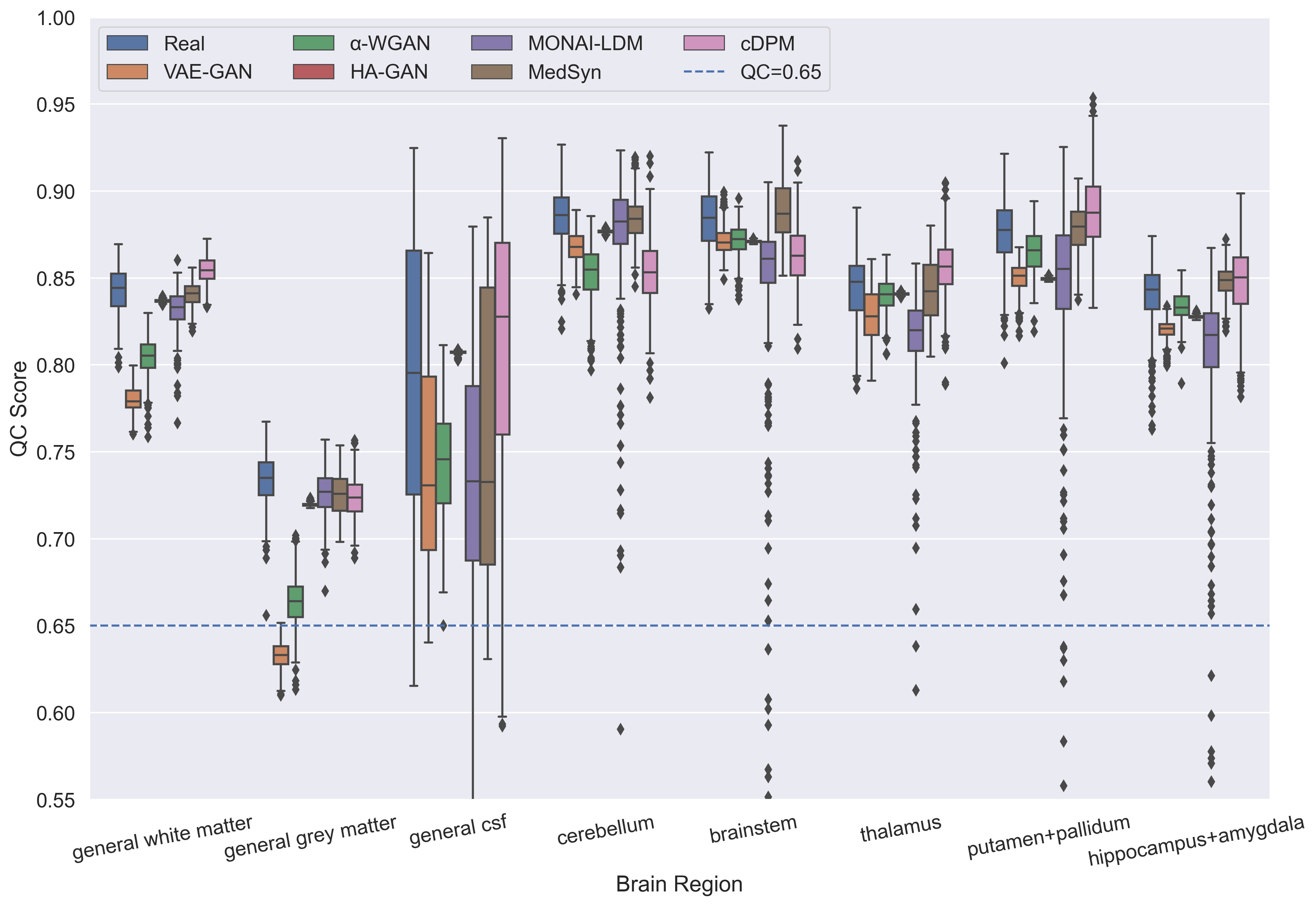}}
    \caption{QC scores of 8 brain regions and 400 MRIs produced by y $Synthseg^{+}$.}
    \label{fig:qc}
\end{figure}

These shortcomings are the motivation behind our framework for computing anatomical plausibility. The first step in computing that metric is to check that the segmentations extracted from the MRIs are reliable. For each method Table \ref{tab:qc} lists the total number of ROI segmentations and corresponding MRIs that failed the check (i.e, QC score < 0.65, which is the case for 4.75\% of real MRI).  Interestingly, $Synthseg^{+}$ produces segmentations that look realistic even if the MRIs are of bad quality as in the case of VAE-GAN (see Fig. \ref{segvisual}). However, the quality score generated by $Synthseg^{+}$ clearly points out that most of those segmentations should not be trusted as less than 1$\%$ pass quality control according to Table \ref{tab:qc}. Performing much better is $\alpha$-WGAN (success rate $83.25\%$) but the quality of its gray matter segmentation often fails to meet the threshold according to the plot shown in Fig. \ref{fig:qc}. A slightly better success rate has MONAI-LDM with $86\%$. However, when an MRI produced by this method fails to meet the QC threshold, it is not for a specific region and often involves multiple ones. Only the three models with the best-looking MRIs pass the 95\% threshold, which are HA-GAN and the diffusion models cDPM and MedSyn.The gray matter is the only region failing QC for few  MRIs generated by cDPM and MedSyn, while HA-GAN is the only model, where all MRIs passed QC. Interestingly enough, the MRIs produced by HA-GAN are visually inferior to those of the two diffusion models so the QC failure rate should not be used as a final metric for assessing anatomical plausibility.  




\subsubsection{QC detection and ROI Comparison}

\begin{table}[!t]
\caption{Rate of MRIs passing QC check}
\label{tab:qc}
\resizebox{\columnwidth}{!}{
\begin{tabular}{@{}lccccccc@{}}
\toprule
                  & real MRIs  & VAE-GAN  & $\alpha$-WGAN & HA-GAN  & MONAI-LDM & MedSyn & cDPM \\ \midrule
Total failed ROI  & 19        & 399      & 67     & 0      & 95      & 6        & 11        \\
Total failed MRIs & 19        & 397      & 67     & 0      & 56      & 6        & 11        \\
MRIs passing rate & 95.25\% & 0.75\% & 83.25\%  & 100.00\% & 86.00\%   & 99.00\%   & 97.25\%   \\ \bottomrule
\end{tabular}}
\end{table}

\begin{table}[!t]
    \newcommand{\fontCohen}{\scriptsize}
    \caption{Cohen's d of Cortical and Subcortical ROIs with best score in bold \vspace{1mm} }
    \begin{minipage}[b]{1.0\linewidth}
    \fontCohen
    \begin{tabular*}{\linewidth}{@{\extracolsep{\fill}} lcccccccc}
           & cerebral & \multicolumn{2}{|c|}{lateral ventricle} & \multicolumn{2}{c|}{cerebellum} & \multicolumn{1}{c|}{thalamus}     & \multicolumn{1}{c|}{caudate}        & putamen  \\
           &   WM    &  \multicolumn{1}{|c}{}                 &  \multicolumn{1}{c|}{inferior} &  WM & \multicolumn{1}{c|}{GM}            &   \multicolumn{1}{c|}{}   & \multicolumn{1}{c|}{}  &  \\ \hline
    HA-GAN  & 0.68                                                  & 0.10                                                     & 0.50                                                     & -0.35                                                    & -0.59          & -0.18         & 0.39           & 0.53           \\
    MedSyn & -0.12                                                 & 0.69                                                     & 0.69                                                     & \textbf{0.22}                                            & 0.26           & -0.39         & -0.31          & -0.17          \\
    cDPM   & \textbf{0.00}                                         & \textbf{-0.05}                                           & \textbf{-0.15}                                           & 0.25                                                     & \textbf{0.23}  & \textbf{0.05} & \textbf{-0.12} & \textbf{-0.08} \\ \hline 
    \multicolumn{9}{l}{~~WM=white matter, GM=gray matter } \vspace{2mm}  \\
           & pallidum  & \multicolumn{2}{|c|}{ventricle}  & brain- & \multicolumn{1}{|c|}{hippo-}  & amygdala      & \multicolumn{1}{|c|}{accumbens}            & cerebrospinal            \\
           &           & \multicolumn{1}{|c}{3$^{rd}$} & \multicolumn{1}{c|}{4$^{th}$}   & stem   & \multicolumn{1}{|c|}{campus}  &   & \multicolumn{1}{|c|}{}& fluid\\ \hline
    HA-GAN  & 1.16                                                  & \textbf{0.09}                                            & 0.92                                                     & -0.34                                                    & -1.01          & -0.04         & 0.68           & 0.31           \\
    MedSyn & -0.38                                                 & 0.15                                                     & 0.34                                                     & -0.41                                                    & -0.79          & -0.76         & -0.21          & -0.51          \\
    cDPM   & \textbf{0.01}                                         & -0.16                                                    & \textbf{0.06}                                            & \textbf{0.18}                                            & \textbf{-0.02} & \textbf{0.02} & \textbf{-0.07} & \textbf{0.04}  \vspace{1.5mm}  \\
    \hline \hline \vspace{-2mm}
    \end{tabular*}
    \label{tab:CohenD}

    \fontCohen
     \begin{tabular*}{\linewidth}{@{\extracolsep{\fill}} lcccccccc}
       & bankssts & \multicolumn{2}{|c|}{caudal} & cuneus & \multicolumn{1}{|c|}{entorhinal} & fusiform & \multicolumn{2}{|c}{inferior}\\ 
       &   & \multicolumn{1}{|c}{AC} & \multicolumn{1}{c|}{MF} &  & \multicolumn{1}{|c|}{} & & \multicolumn{1}{|c}{parietal} & temporal \\
       \hline
HA-GAN & 0.25 & 1.13 & 0.26 & 1.09 & 1.11 & 2.00 & 1.08 & \textbf{0.06} \\
MedSyn & 0.26 & 0.56 & -0.38 & 0.27 & 0.95 & \textbf{-0.11} & -0.09 & -0.25 \\
cDPM & \textbf{0.04} & \textbf{-0.21} & \textbf{-0.09} & \textbf{0.11} & \textbf{0.07} & 0.28 & \textbf{0.00} & 0.32 \\ \hline
\vspace{1mm} \\ 
& \multicolumn{1}{c|}{isthmus} & \multicolumn{2}{c|}{lateral} & \multicolumn{1}{c|}{frontal-} & \multicolumn{1}{c|}{lingual} & medial & \multicolumn{1}{|c|}{middle} & \multicolumn{1}{c}{parahippo-}\\ 

& \multicolumn{1}{c|}{cingulate}   & \multicolumn{1}{c}{occipital} & \multicolumn{1}{c|}{orbitofrontal}  & \multicolumn{1}{c|}{pole} & \multicolumn{1}{c|}{}&\multicolumn{1}{c}{orbitofrontal} &\multicolumn{1}{|c|}{temporal}& \multicolumn{1}{c}{campal}\\
       \hline
HA-GAN & 1.32 & 0.93 & -0.28 & 1.12 & 0.24 & 0.47 & -0.38 & \textbf{-0.09}  \\
MedSyn & \textbf{0.07} & -0.41 & -0.43 & -0.35 & -0.72 & -0.47 & -0.12 & -0.14 \\
cDPM & 0.11 & \textbf{0.15} & \textbf{-0.06} & \textbf{-0.07} & \textbf{0.22} & \textbf{0.04} & \textbf{0.01} & 0.13 \\\hline
\vspace{0mm} \\ 
& \multicolumn{3}{c|}{central} & \multicolumn{3}{c|}{pars} & \multicolumn{1}{c|}{peri-} & temporal-  \\ 
& pre & para & \multicolumn{1}{c|}{post}  & \multicolumn{1}{c}{opercularis} & \multicolumn{1}{c}{orbitalis}  & \multicolumn{1}{c|}{triangularis} &\multicolumn{1}{c|}{calcarine} & pole \\
       \hline
HA-GAN &  -0.26 & -1.10 & -1.81  & 0.92 & 0.73 & -0.16 & -0.58 & -1.00 \\
MedSyn &  0.40 & 0.32 & 0.38 &  -0.31 & 0.16 & 0.27 & -1.37 & \textbf{0.04}  \\
cDPM   & \textbf{-0.02}  & \textbf{0.02} & \textbf{-0.02} & \textbf{-0.10} & \textbf{0.00} & \textbf{-0.05} & \textbf{0.03} & 0.10 \\ \hline 
\vspace{0.2mm} 
\end{tabular*}
%
\fontCohen
 \begin{tabular*}{\linewidth}{@{\extracolsep{\fill}} lccccccccc}
& \multicolumn{1}{c|}{cingulate} & precuneus & \multicolumn{2}{|c|}{rostral} & \multicolumn{3}{c}{superior-} & \multicolumn{1}{|c|}{supra-} & \multicolumn{1}{c}{transverse}  \\ 

& \multicolumn{1}{c|}{}   &    & \multicolumn{1}{|c}{AC}  & \multicolumn{1}{c|}{MF} &\multicolumn{1}{c}{frontal} &\multicolumn{1}{c}{parietal}& temporal & \multicolumn{1}{|c|}{marginal}& temporal\\
       \hline       
HA-GAN &  -0.22   & -0.41  & 1.13 & 1.02 & 0.14 & 1.18  & -0.58 & -0.95 & 0.17\\
MedSyn & \textbf{-0.04} & -0.61 & \textbf{0.11} & 0.42 & \textbf{-0.12} & \textbf{0.14} & \textbf{-0.07} & 0.17 & 0.14  \\
cDPM  & 0.09  & \textbf{0.04} & -0.16  & \textbf{-0.18} & -0.28 & 0.21 & -0.08  & \textbf{0.00}   & \textbf{0.08} \\ \hline
\multicolumn{9}{l}{~~AC=anterior cingulate, MF=middle frontal}
\end{tabular*}

\end{minipage}
\end{table}
The anatomical plausibility of the MRIs produced by those three methods is summarized by the Cohen’s d score between the volume distributions of real and synthetic MRIs for each of the 16 subcortical regions and 33 cortical regions (Table \ref{tab:CohenD}). A value closer to 0 indicates higher overlap between the distributions and thus anatomical plausibility. Reflecting visual assessment,  diffusion models generally produce MRIs of higher anatomical plausibility than HA-GAN, as cDPM has the best Cohen’s d score for 38 regions (i.e., 77.5$\%$ of regions), followed by MedSyn with 9 regions (18.4$\%$), and HA-GAN with 2 regions (4.1$\%$). Note, that one can also use the anatomical plausibility scores proposed here to exclude the MRIs for aiding the analyzes focusing on regions in which these MRIs receive poor scores (e.g., $|d|$>0.8), such as recorded for 14 regions by HA-GAN. 

\section{Conclusion}
This work not only documents the shortcomings in current assessments of structural brain MRIs produced by generative models but also proposes a framework for solving these issues. The framework standardizes the experimental setup for comparing methods and assessing anatomical plausibility, a metric we previously introduced in \cite{peng2023metadataconditioned}. Unlike in \cite{peng2023metadataconditioned}, we ensure that the metric returns reliable results by checking that MRI segmentations meet a pre-defined quality threshold. We use this framework to compare six state-of-the-art methods revealing that diffusion models generally produce higher-quality MRIs than generative adversarial networks. More importantly, these assessments align with the visual quality of the MRIs displayed in this article. By creating a reliable assessment for generated MRIs, this framework provides a critical step toward using these images to advance MRI studies.

\begin{credits}
\subsubsection{\ackname} Part of the data set used for this analysis and work was supported by funding from the National Institute of Health (DA057567, AA021681, AA021690, AA021691, AA021692, AA021695, AA021696, AA021697, AA017347, AA010723, AA005965, and AA028840), the DGIST R\&D program of the Ministry of Science and ICT of KOREA (22-KUJoint-02), and the Stanford HAI Google Cloud Credits.
\subsubsection{\discintname} The authors have no competing interests to declare that are relevant to the content of this article.

\end{credits}

\bibliographystyle{splncs04}
\bibliography{Paper-0689}

\end{document}